\documentclass[sigconf]{acmart}

\AtBeginDocument{%
  \providecommand\BibTeX{{%
    \normalfont B\kern-0.5em{\scshape i\kern-0.25em b}\kern-0.8em\TeX}}}

    \usepackage{enumitem}
\usepackage{multirow}
\usepackage{xcolor}
\usepackage{adjustbox,lipsum}
\usepackage{subcaption}
\usepackage{blindtext}
\usepackage{array}
\usepackage{nameref}
\usepackage{footnote}
\usepackage{balance}
\usepackage{rotating}
\usepackage{tablefootnote}
\usepackage{hyperref}
\usepackage{dirtytalk}
\makesavenoteenv{tabular}
\makesavenoteenv{table}
\usepackage{geometry}
\usepackage{pdflscape}
\usepackage{orcidlink}

\newcommand\changed[1]{\textcolor{blue}{#1}}

\copyrightyear{2024}
\acmYear{2024}
\setcopyright{rightsretained}
\acmConference[ESEM '24]{Proceedings of the 18th ACM / IEEE International Symposium on Empirical Software Engineering and Measurement}{October 24--25, 2024}{Barcelona, Spain}
\acmBooktitle{Proceedings of the 18th ACM / IEEE International Symposium on Empirical Software Engineering and Measurement (ESEM '24), October 24--25, 2024, Barcelona, Spain}\acmDOI{10.1145/3674805.3686687}
\acmISBN{979-8-4007-1047-6/24/10}

\begin{document}
\title{Understanding Fairness in Software Engineering: Insights from Stack Exchange Sites}

\author{Emeralda Sesari}
\affiliation{%
  \institution{University of Groningen}
  \city{Groningen}
  \country{Netherlands}
  }
\email{e.g.sesari@rug.nl}

\author{Federica Sarro}
\affiliation{%
  \institution{University College London}
  \city{London}
  \country{United Kingdom}
  }
\email{f.sarro@ucl.ac.uk}

\author{Ayushi Rastogi}
\affiliation{%
  \institution{University of Groningen}
  \city{Groningen}
  \country{Netherlands}
  }
\email{a.rastogi@rug.nl}

\renewcommand{\shortauthors}{Emeralda Sesari, Federica Sarro, Ayushi Rastogi}

\begin{abstract}
Software practitioners often discuss technical and social workplace issues, both in-person and online.
One such social issue is fairness. As research on fairness in software engineering expands, it often concentrates on specific issues.
This exploratory study provides discussions on Stack Exchange sites, focusing on the experiences and expectations of fairness in software engineering.
We also want to identify the fairness aspects software practitioners talk about the most. 
For example, do they care more about fairness in income or how they are treated in the workplace?

Our investigation of fairness discussions on eight Stack Exchange sites resulted in a list of 136 posts (28 questions and 108 answers) manually curated from 4,178 candidate posts.
The study reveals that the majority of fairness discussions 
revolve around the topic of income suggesting that many software practitioners are highly interested in matters related to their pay and how it is fairly distributed.
Further, we noted that while not discussed as often, discussions on fairness in recruitment tend to receive the highest number of views and scores. Interestingly, the study shows that unfairness experiences extend beyond the protected attributes
, with gender mainly being discussed.
\end{abstract}




\maketitle

\section{Introduction}
Software practitioners encounter a range of problems at work.
Some of these problems relate to their work, for example, incomplete requirements, software testing, and code-related problems \cite{Ragkhitwetsagul2022}.
Other problems relate to human and social aspects, for example, personality clashes inhibiting collaboration~\cite{Capretz2014} and the need for improved customer relations and communications~\cite{Lenberg2015}. 
One such factor is fairness at work.
Studies show that understanding fairness problems can improve software products that can indirectly impact software companies' profit \cite{Capretz2014}. 
German et al. \cite{German2018} developed a framework to study the perceptions of fairness in modern code reviews. 
Another study found differences between male and female practitioners when expressing sentiments during Software Engineering (SE) interactions and how their opinions differ based on the gender of the recipient indicating gender bias~\cite{Paul2019}. 
Further, there are studies investigating the challenges women face in global software development teams~\cite{Trinkenreich2022} and in open source software development~\cite{Imtiaz2019}. 


While previous studies \cite{German2018, Paul2019, Trinkenreich2022, Imtiaz2019, Gunawardena2022DestructiveCI} have focused on specific issues related to fairness in SE, there remains a broader gap in the literature. These studies highlight the potential for improving developer satisfaction and productivity through more fair work practices. Our study aims to explore the overall concept of fairness and understand the specific fairness-related concerns of software practitioners, providing a comprehensive view that builds on these previous insights.
We align our understanding of fairness relating to human and social factors in SE by building upon the theoretical framework provided by Organizational Justice (OJ) literature \cite{Colquitt2001}, a framework that has previously informed fairness exploration in SE work by German et al. \cite{German2018}. Fairness refers to the perception that the practices, policies, and interactions within an organization are unbiased, and just \cite{Colquitt2001}. Fairness relating to human and social factors in SE encompass, for example, fair task allocation, respectful interactions, and clear communication within engineering teams to promote a fair and supportive work environment.


This paper explores fairness discussions among software practitioners on \textsc{Stack Exchange} (\textsc{StEx}) sites. \textsc{StEx}, a network of question-and-answer websites, offers valuable insights into practitioners' behaviors and perspectives, which are often difficult to gather through surveys or interviews due to potential biases. Fairness have an impact on various aspects, from algorithms \cite{Brun2018} to team dynamics. The first research question explores the usefulness of \textsc{StEx} sites for analyzing fairness discussions relating to human and social factors in SE. 
To gauge the usefulness of \textsc{\textsc{StEx}} sites, we inquire:

\textbf{RQ1: How often is fairness, relating to human and social factors in SE, discussed on \textsc{StEx} sites?} 
      \label{itm:rq1}
We analyzed eight \textsc{\textsc{StEx}} sites likely to discuss fairness in human and social factors in SE. 
This includes \textsc{Stack Overflow}, \textsc{Software Engineering \textsc{StEx}}, \textsc{Workplace \textsc{StEx}}, \textsc{Code Review \textsc{StEx}}, \textsc{Open Source \textsc{StEx}}, \textsc{Computer Science \textsc{StEx}}, \textsc{DevOps \textsc{StEx}}, and \textsc{Project Management \textsc{StEx}}. 
We identified 1,354 candidate fairness posts across five \textsc{StEx} sites, of which 136 posts (10.0\%) relate to fairness in human and social factors in SE. 
This semi-automated process employs a mixed-methods empirical approach, consisting of keyword searches and manual analysis.
RQ1 reveals that the \textsc{StEx} sites have various discussions on fairness, covering areas such as algorithms, computation, and human and social factors. We find that 10 out of every 100 candidate fairness posts relate to human and social aspects of SE, with most of them being hosted on \textsc{Workplace StEx}.
The second research question explores the fairness problems discussed by software practitioners and their characteristics.

\textbf{RQ2: What characterizes fairness discussions on \textsc{StEx} sites?}
We characterize fairness discussions on three attributes. 

  \emph{RQ2.1: What are the contexts in which fairness is discussed?}  
We manually identified 10 specific fairness contexts and categorized the fairness posts accordingly. The results indicate that a majority of posts (24 out of 136 posts) discuss fairness in income. Then, we map the selected posts to the fairness dimensions borrowed from OJ literature.

    \emph{RQ2.2: Which dimensions of fairness are discussed?} To this end, we employed Colquitt's fairness measure \cite{Colquitt2001}, widely used in OJ literature \cite{German2018}. 
    Our study showed that practitioners are most concerned about fairness in decision-making processes influencing the outcome (53 fairness posts or 38.97\%), consistent with what has been observed generally in OJ literature.
    Last, we investigate the demographics of the post owners for characterization in order to understand the potential relationship between user characteristics and the fairness posts.

        \emph{RQ2.3: What are the characteristics of the post owners?} We inferred post owners' demographics using manual analysis.
        Out of 14 inferable users' countries, the United States accounts for 50.67\% users. We observed that the distribution of users writing fairness posts in \textsc{Workplace StEx}, \textsc{Software Engineering StEx}, and \textsc{Project Management StEx} relates to the sites' population distribution.

The key contributions of our work include the following:
\begin{itemize}[noitemsep,topsep=2pt]
\item  knowledge on fairness contexts and dimensions that software practitioners care about through the first exploration conducted on fairness questions and answers from \textsc{StEx} posts;  

\item characterizing posts about fairness in software engineering relating to the fairness dimensions described in the OJ literature;

\item a set of unique contexts in which fairness in software engineering is discussed, accompanied by an analysis of the demographics of the post owners; 

\item a manually annotated set of fairness posts useful for building automated tools and extending this work by focusing on specific fairness contexts or dimensions. 
\end{itemize}
Our online repository \cite{figshare} contains source code used to retrieve and analyze data, results for all our RQs, and instructions how to replicate the results. The study has been concluded to be ethically appropriate by our institution.

\section{Background and Related Work}
\label{sec:background}
\subsection{Mining Stack Exchange Posts} 
\textsc{Stack Exchange (\textsc{StEx})} is a network of 181 communities that are created and run by experts and enthusiasts who are passionate about a specific topic \cite{SEintro}. 
\textsc{Stack Overflow (SO)}, an online platform found in 2008 where programmers could help answer each other problems for free, was the origin of \textsc{StEx}. Ever since, the network has expanded to include many other communities, such as technology (e.g.\changed{,} \textsc{Game Development}), professional (e.g., \textsc{Workplace}), and business (e.g., \textsc{Quantitative Finance}).
On \textsc{StEx} sites, a user can ask \textbf{questions} and receive \textbf{answers} from other users. Good answers are \textbf{voted up} and shown first, making them easy to locate. The owner of the question can \textbf{mark} one answer as "accepted," indicating that it worked for the person who asked. 
\textbf{Comments} can be used to request additional information or to clarify a question or answer.

\textsc{StEx} sites provide a rich source of information, offering insights into understanding practitioners' behaviors, interactions, and perspectives on specific themes \cite{Treude2011}. Many studies have used \textsc{SO} posts, comments, tags, and user profiles to gain insight into important topics. For example, Khristul et al. investigated human values violations in \textsc{SO} posts, finding that 315 out of 2000 random comments expressed concerns that linked posts contradict human values \cite{khristul2022}. Ferreyra et al. \cite{Ferreyra2022} studied developers’ self-disclosure behavior on \textsc{SO}, revealing that proactive users provide much less information in their profiles than reactive and unengaged individuals. These studies demonstrate that \textsc{SO} is a valuable data source for uncovering useful information about software developers’ behaviour and perspective.


In this study, we are interested in the fairness problems software practitioners experience in software engineering. Practitioners place great trust in their peers, particularly local experts \cite{Rainer2003, Devanbu2016}. Therefore, software practitioners are more likely to rely on sources of knowledge they trust, such as \textsc{StEx} sites (including \textsc{SO}) which is referred to as Grey Literature (GL) \cite{benzies2006}. Researchers, on the other hand, often place great trust in peer-reviewed, published knowledge, such as journals and conference articles \cite{Kratz2015}. The difference in sources of knowledge might result in misunderstanding where researchers and practitioners have different perceptions of what is relevant \cite{Garousi2017}.
We intend to fill this gap by looking into practitioners' discussions about fairness in SE on \textsc{StEx} sites and understand what aspects of fairness practitioners care about.

\subsection{Fairness in Software Engineering}
Software is a byproduct of problem-solving, cognitive characteristics, and social interaction \cite{Capretz2014}. Among the various activities, social activities can be a key differentiator for most software projects. In SE, management views knowledge of social interaction as crucial \cite{Capretz2014}. Managers, for example, must learn about human factors since they constantly deal with team member negotiations and personality conflicts \cite{Capretz2014}. 
Managers also recognize the need to recruit skilled software practitioners for the correct position. 
Since fairness is a fundamental element to keep cooperative relationships with others \cite{Dagger2018Fair}, understanding fairness problems may result in improved software products, increased team efficiency, and lower software production costs, all of which may contribute to improving the software company's profit \cite{Capretz2014}.

Several studies investigate how bias and unfairness prevail. Trinkenreich et al. \cite{Trinkenreich2022AnEI} surveyed Ericsson software developers and found that socio-cultural barriers prevent women from advancing in global software development teams. Survey participants indicate that women's voices are silenced by men and they are included in projects based on gender rather than ability \cite{Trinkenreich2022AnEI}. In code review, Paul et al. \cite{Paul2019} found that male developers in three out of six projects were writing more negative comments and refusing to assist their female peers, leading to disparities in sentiment expressions between male and female practitioners. Similarly, Gunawardena et al. \cite{Gunawardena2022DestructiveCI}  discovered that destructive critique in code reviews affects participants' moods, desire to continue working, and written reactions to the criticism. Women are found to perceive destructive criticism as less acceptable and are less motivated to work with the developer after receiving it. Therefore, destructive criticism in code reviews may contribute to the lack of gender diversity observed in the software industry.

The closest to our study is that of German et al., who employed Colquitt's fairness theory \cite{Colquitt2001, Colquitt2002} to develop a framework explaining fairness in the modern code review process \cite{German2018}. 
The framework can be used to examine code review processes to identify strengths and potential areas for improvement. 
For example, a system that supports equity, one of the rules under distributive fairness \cite{Colquitt2001}, assumes that the patch's author and the qualities of the patch should influence resource allocation and treatment. The authors also explore the developers' concerns about fairness in code review by surveying developers practising the OpenStack code review process \cite{German2018}. The responses indicate that several contributors are concerned about unfairness, particularly equity, equality, consistency, and bias suppression \cite{German2018}. 
Our study, instead, explore the fairness problems that practitioners experience in the SE practice reflected in practitioners' posts on \textsc{StEx} sites. We refer to Colquitt's fairness theory to categorize the posts into the fairness dimensions that best represent the fairness problem discussed. This paper provides new insights to the research presented by German et al. \cite{German2018} by identifying fairness problems in a SE context beyond the code review process. This prior research by German et al. \cite{German2018} underscores the importance and feasibility of applying fairness measure in SE, supporting our approach to extend these concepts to broader SE practices.



\subsection{Fairness in Organizational Justice}
Organizational Justice (OJ) refers to actions taken within the organization based on what is considered morally right and wrong \cite{Colquitt2012}. 
For four decades, fairness has been a widely researched area for scholars interested in OJ \cite{Colquitt2013}. 
Justice and fairness, according to Colquitt et al. \cite{Colquitt2015} and Cuguero et al. \cite{Cuguero2013}, are similar but different notions. Fairness should be understood to refer to a subjective assessment or evaluation of organizational activities and if the activities, as they are being carried out, are moral and praiseworthy \cite{Goldman2015}. Justice, on the other hand, refers to the rules and procedures that are put in place to ensure that fairness is achieved \cite{Colquitt2015}.

In the early years of justice literature, various fairness dimensions have been discovered. Initially, justice scholars concentrated on the fairness of decision outcomes, known as \textbf{distributive fairness} \cite{ADAMS1965267,Deutsch1975,LEVENTHAL197691}. 
It is promoted when outcomes are aligned with implicit allocation rules such as equity or equality \cite{ADAMS1965267,Deutsch1975,LEVENTHAL197691}. Not long after that, multiple studies focused on the fairness of the procedures that lead to decision outcomes, namely \textbf{procedural fairness} \cite{Leventhal1980,thibaut_walker_1975, leventhal_karuza_1980}. 
It is promoted through participation in decision-making or influence over the outcome \cite{thibaut_walker_1975}, as well as adherence to fair process standards such as consistency, lack of bias, correctability, representation, accuracy, and ethicality \cite{Leventhal1980, leventhal_karuza_1980}. In 1993, Greenberg \cite{Greenberg1993} proposed the components of distributive fairness that involve respect and sensitivity may best be understood as \textbf{interpersonal fairness} since they influence reactions to decision outcomes. The author further recommended that the part of procedural fairness that involves providing clear and accurate explanations to those affected by a decision may best be considered as \textbf{informational fairness} because explanations frequently give the information required to evaluate the structures of the procedure.
In 2001, Colquitt \cite{Colquitt2001} developed a list of questions (see Table \ref{tab:dimensions}) to describe the fairness dimension that can be used to assess the fairness of a situation in various contexts, including management \cite{Roberson2012} and psychology \cite{Lis2018}. 

\begin{table*}[]
\captionsetup{font=footnotesize}
\footnotesize
\caption{Fairness Dimensions and Colquitt's questions to identify them \cite{Colquitt2001}}
\label{tab:dimensions}
\begin{adjustbox}{max width=\linewidth}
\begin{tabular}{ll}
\multicolumn{1}{l}{\textbf{Dimensions}} & \multicolumn{1}{l}{\textbf{Questions}}                                                                                                           \\ \hline
\multirow{4}{*}{Distributive}   & Does your outcome reflect the effort you have put into your work?                                                                        \\
                                & Is your outcome appropriate for the work you have completed?                                                                              \\
                                & Does your outcome reflect what you have contributed to the organization?                                                                  \\
                                & Is your outcome justified, given your performance?                                                                                        \\ \hline
\multirow{7}{*}{Procedural}     & Have you been able to express your views and feelings during those procedures?                                                                \\
                                & Have you had influence over the outcome arrived at by those procedures?                                                                        \\
                                & Have those procedures been applied consistently?                                                                                                \\
                                & Have those procedures been free of bias?                                                                                                        \\
                                & Have those procedures been based on accurate information?                                                                                       \\
                                & Have you been able to appeal the outcome arrived by those procedures?                                                     \\
                                & Have those procedures upheld ethical and moral standards?                                                                                       \\ \hline
\multirow{4}{*}{Interpersonal}  & Has (the authority figure who enacted the procedure) treated you in a polite manner?                                                        \\
                                & Has (the authority figure who enacted the procedure) treated you with dignity? \\
                                & Has (the authority figure who enacted the procedure) treated you with respect?                                                              \\
                                & Has (the authority figure who enacted the procedure) refrained from improper remarks or comments?                                           \\ \hline
\multirow{5}{*}{Informational}  & Has (the authority figure who enacted the procedure) been candid in their communications with you?                     \\
                                & Has (the authority figure who enacted the procedure) explained the procedures thoroughly?                                                   \\
                                & Were (the authority figure who enacted the procedure)’s explanations regarding the procedures reasonable?                                   \\
                                & Has (the authority figure who enacted the procedure) communicated details in a timely manner?                                               \\
                                & Has (the authority figure who enacted the procedure) seemed to tailor their communications to individual’s specific needs?                  \\ \hline
\end{tabular}
\end{adjustbox}
\end{table*}

\section{Methodology}
\label{methodology}
We provide step-by-step description on how we gathered and analyzed practitioners' perspectives on fairness using \textsc{StEx} posts. First, we selected \textsc{StEx} sites that were likely to be useful for our search. We then identified candidate fairness posts by conducting a keyword search followed by manually analyzing the posts to identify fairness posts relating to human and social aspects. We repeated the same search with additional keywords to find fairness posts. We tried to mitigate researcher's bias by applying consensus coding on the manual extraction process.

\textbf{Step 1: Identify sites.}
The goal in this study is to understand the fairness problems that software practitioners face during software development. Gaining insights into these problems can be difficult since they often involve sensitive information based on personal experiences. \textsc{StEx} sites, which host discussions on a wide range of topics, offer valuable data, including conversations about fairness in SE. We believe \textsc{StEx}'s long standing role in the development community makes it an ideal starting point for our study. The anonymity provided to users encourages candid discussions about their experiences \cite{Alnasser2021An} making the information a direct reflection of the experiences faced by users.

While searching all \textsc{StEx} sites will be comprehensive, it can also be overwhelming and time-consuming. We believe some sites may not be relevant to the study, making it efficient to focus on the most relevant sites. For example, we did not focus on \textsc{Mathematics StEx}, which caters to people studying math, and \textsc{Arqade StEx}, which caters to people playing video game.
From 181 \textsc{StEx} sites, we selected posts from 8 sites that could be related to SE and fairness. Please refer to Table \ref{tab:sitestopics} to understand the justification behind selecting each site. We recognize that conversations related to SE and fairness may take place across various StEx sites such as \textsc{Software Quality Assurance \& Testing StEx}. However, to maintain focus and manageability of our dataset, we concentrated our data collection efforts on sites that we believed were most likely to provide valuable insights into our research questions. Next, we downloaded the public release \cite{datadump} of an anonymized dump of all user-contributed content on all 8 sites with last modified dates between 3 and 9 October 2022. 
All data dump of 181 \textsc{\textsc{StEx}} community sites has been made publicly available \cite{datadump}.

\begin{table}[]
\footnotesize
\captionsetup{font=footnotesize}
\caption{Selected \textsc{StEx} community sites}
\label{tab:sitestopics}
\begin{adjustbox}{max width=\columnwidth}
\begin{tabular}{p{.3\linewidth}p{.7\linewidth}}
\multicolumn{1}{c}{\textbf{Community Sites}} & \multicolumn{1}{c}{\textbf{Justification}}                                                                                                                                                                                                                                                                                                \\ \hline
\multirow{2}{8em}{Stack Overflow (SO) }                     &  Popular with the SE community and  focuses \cite{StackOverflow1} is on problems specific to software development  \cite{StackOverflow1}                                                                                                         \\ \hline
\multirow{4}{8em}{Software Engineering (SE) \textsc{StEx}     }        & The site covers: software development methods and practices; requirements, architecture, and design; quality assurance and testing; configuration management; build, release, and deployment  \cite{SoftwareEngineeringSE}                                                                                               \\ \hline
\multirow{2}{8em}{Workplace (WP) \textsc{StEx}} & Fairness is usually assessed in the context of organization such as in the workplace \cite{greenberg2011organizational}                                                                                                                                                                             \\ \hline
\multirow{3}{8em}{Code Review (CR) \textsc{StEx}} & Issues of fairness in code review include how it is done, the resources it gets, and how the reviewers treat the person submitting the patch \cite{German2018}                                                                                                              \\ \hline
\multirow{3}{8em}{Open Source (OS) \textsc{StEx}}& In OS software projects, people expect to receive fair treatment even if they do not all have the same idea of what is fair \cite{German2018}                         \\ \hline
\multirow{2}{8em}{Computer Science (CS) \textsc{StEx}}& The site may discuss aspects of software and software systems in the design and development phases \cite{CSSE}                                                                                              \\ \hline
\multirow{3}{8em}{DevOps (DO) \textsc{StEx} }& The site covers automated testing, continuous delivery, service integration and monitoring, Software Development Life Cycle (SDLC) \cite{DevOpsSE}                                                                                                                                \\ \hline
\multirow{3}{8em}{Project Management (PM) \textsc{StEx} }& Managing a software project involves working with many people over a long period of time, and it affects how different groups of people work and perform \cite{Boehm1989}\\ \hline
\end{tabular}
\end{adjustbox}
\end{table}

\textbf{Step 2: Identify fairness posts.}
The posts in the selected 8 \textsc{StEx} sites can be about any topic within the scope of SE whereas we needed to identify fairness posts only. Previous studies used keywords to collect data such as \textsc{SO} posts \cite{khristul2022} and user reviews \cite{Obie2020} containing human values violations.

\textit{Keyword Search.} We first used the term \textit{fair} to find posts about fairness. The term \textit{fair} is a direct and concise word that often represents the essence of fairness, making it a suitable starting point for our search. Our analysis of a random sample of posts showed that, the resulting posts also include posts that are not necessarily about fairness. For instance, \textit{to be fair}, \textit{fairly}, \textit{fair enough}, \textit{fair catch}, \textit{job fair}, and \textit{fair to say}. Therefore, for the first filtering, we selected the posts by running a query using the term \textit{fairness} to minimize false positives in comparison to using the term \textit{fair}. We chose to consider both types of posts, namely question and answer for two reasons. First, a question can communicate a user's concern, while an answer can provide a solution which addresses that concern. Column \textit{1st Filtering} in Table \ref{tab:extraction} shows the number of candidate posts after filtering.


\textit{Manual Extraction}. We manually analyzed the filtered posts to identify posts related to human and social factors in SE. 
 Previous study shows that finding insights hidden in qualitative data requires consistency \cite{Basit2003}. For a small set of data, manual analysis can be effective in maintaining that consistency. Previous studies also conducted manual extraction on \textsc{SO} posts \cite{khristul2022, Obie2020}. 
In summary, we manually extract posts that have relevance to fairness in human and social aspects of SE using the following criteria:

\begin{itemize}[noitemsep,topsep=2pt]
    \item [IC1:] the post contains at least one explicitly described fairness problem, or,
    \item [IC2:] the post implicitly addresses the problem of fairness through phrases such as \say{is this appropriate?};
    \item [IC3:] the activities include software practitioners (e.g., engineers, developers, designers), software lifecycle processes, or organizational policies.
\end{itemize}

Exclusion criteria:
\begin{itemize}[noitemsep,topsep=2pt]
    \item [EC:] Fairness in computation paths, modern software or algorithm.
\end{itemize}

Fairness can be described explicitly, such as in phrases \say{recognize the fairness of your responses} and \say{for the sake of fairness}. It can also be described implicitly by stating the user's perception of appropriateness \cite{Colquitt2015AddingT} such as in phrases \say{is this appropriate?} and \say{what is the right thing to do?}. Therefore, we include this implicit notion of fairness in addition to the posts that explicitly describe fairness (IC1 and IC2). Lastly, the fairness problem must take place in a software development setting (IC3). Fairness post must include the people who are involved in SE process namely software practitioners; the activities surrounding software creation (i.e., software lifecycle processes); and the managerial or organizational policies surrounding software creation (e.g., hiring practitioners, allocating tasks). 
By adhering to these strict criteria, we sought to ensure consistency of analysis. Unlike automated methods commonly applied to textual data \cite{Barberá2020Automated} such as \textsc{StEx} posts, we emphasized manual extraction. This approach allowed us to capture the detailed aspects of fairness, which automated processes may overlook, resulting in a more comprehensive understanding of fairness in SE practice.

During manual extraction, we discarded posts discussing fairness in computation paths, such as multithreading, multi-processing, and concurrency (EC). Fairness, in this context, refers to the probability that different computation paths can improve what they are working on. We excluded this topic because our interest lies only in the human and social aspects of SE and not in the technical aspects. Additionally, posts about modern software fairness (e.g., developing a fair online multiplayer game) were also omitted because this area falls under the scope of fairness of learning-based systems as opposed to fairness in human and social aspects of SE. After the first filtering, we extracted 74 fairness posts from 2,318 posts through manual extraction, as shown in Table \ref{tab:extraction} Column 3.

\begin{table}[]
\captionsetup{font=footnotesize}
\caption{Number of posts extracted from eight \textsc{StEx} sites}
\label{tab:extraction}
\begin{adjustbox}{max width=\columnwidth}
\begin{tabular}{lrrrrr}
Sites & 1st Filtering & 1st Extraction & 2nd Query & 2nd Filtering & 2nd Extraction \\ \hline
WP                                                                                                                    & 298                                                       & 62 &   3,170                                                      & 300                                                        & 25                                                          \\
SE                                                                                                         & 69                                                        & 10    & 1,570                                                     & 300                                                        & 12                                                          \\
PM                                                                                                            & 18                                                        & 2    & 283                                                      & 283                                                        & 16                                                          \\
OS                                                                                                                   & 9                                                         & 0        & 50                                                  & 50                                                         & 8                                                           \\
DO                                                                                                                        & 0                                                         & 0       & 27                                                   & 27                                                         & 1                                                           \\
SO                                                                                                            & 1,825                                                     & 0      & 112,470                                                    & 300                                                        & 0                                                           \\
CR                                                                                                                  & 51                                                        & 0        & 2,069                                                  & 300                                                        & 0                                                           \\
CS                                                                                                              & 48                                                        & 0       & 1,921                                                   & 300                                                        & 0                                                           \\
 \hline
Total                                                                                                            & 2,318                                                      & 74    &     121,560                                                & 1,860                                                       & 62                                                          \\ \hline
\end{tabular}
\end{adjustbox}
\end{table}

\textbf{Step 3: Identify additional posts.}
As described in the previous step, we started by using the term \textit{fairness} to identify relevant posts. However, while manually extracting data, we realized that there could be more keywords that could lead us to posts related to fairness. Therefore, in this step, we created a new query to find additional posts that may involve discussions about fairness in SE. 

\textit{Keyword Search.} We compiled a list of keywords and phrases likely related to concerns about fairness, such as \textit{bias}, \textit{justice}, \textit{equality}, and \textit{equity}. Inspired by similar studies \cite{khristul2022, Obie2020}, the keywords were also chosen based on our observations from the 74 posts, as well as a list of Colquitt's fairness rules (e.g., equity, respect, bias suppression) \cite{Colquitt2001, Colquitt2002} and their associated synonyms and antonyms. We used entries from the online Thesaurus \cite{thesaurus} and Merriam-Webster \cite{merriam} dictionary 
to identify the associated synonyms and antonyms. 
To improve the search and deal with coherency between words, we only selected semantically and contextually linked words. For example, the word \textit{right} is synonym for \textit{equity}. However, the two words are not always used in the same context or with the same intended meaning, which can increase false positive rate. Therefore, we omitted \textit{right} in our keywords list. The second query yielded 121,560 posts in total, excluding the posts found in first extraction.

\textit{Manual Extraction.} 
To answer our RQs, we must identify if each post meets the criteria for a fairness post. If it does (RQ1), we further identified the context (RQ2.1), the fairness dimension (RQ2.2), and the user's characteristics (RQ2.3). At times, to characterize a post, we had to refer to the question associated with an answer, which will be explained in more detail later. On average, this process took about 5 minutes per post, given an average post size of approximately 261 words. That would make a total of 10,130 hours ($\sim$1,266 working days) if we were to analyze all 121,560 posts.

For feasibility, we took a random sample of 300 posts from each \textsc{StEx} site among those returned by the second query, resulting in a combined sample of 1,860 posts. We anticipated that this would take 387.5 hours ($\sim$48 working days). Note that some filtering of sites, \textsc{OS} and \textsc{DO StEx}, generated less than 300 posts; therefore, we decided to keep all posts. Finally, we applied the same manual extraction process used in the first extraction to 1,860 sample posts. 

\textbf{Step 4: Consensus coding process.}
While scientific research must be conducted objectively, the subjective nature of qualitative data analysis can create bias in the researcher's interpretation. 
To enhance the reliability of the manual extraction process and reduce bias, we engaged all three authors with consistent interpretations reflecting the validity of the findings \cite{fleiss1981measurement}. After the extraction of 2,318 filtered posts and second extraction of 1,860 filtered posts, the first author identified 135 fairness posts from 5 sites and no fairness posts from 3 other \textsc{\textsc{StEx}} sites. 

We conducted a consensus coding process \cite{cascio2019team} to ensure that three authors have a shared understanding of the inclusion and exclusion criteria and agreement on how fairness post is extracted, such that if anybody else repeats the same process, they are likely to find the same result. During this consensus coding, the three authors achieved a perfect agreement. 
In the initial data extraction, the first author identified 135 fairness posts. The second and third authors were given access to a subset (50 posts) of the filtered posts for manual analysis. The same subset of posts were then compared with the extraction done by the first author.
Disagreements arose in situations where additional information was required, such as parent posts. However, once the necessary information was provided, disagreements were promptly resolved. Although one post remained disputed by the first and second authors even after receiving additional information, it was reviewed with the third author. Finally, all authors agreed to include this one post, and identified 8 of the 50 posts to be relevant in total.

Given our consensus on the relevance of the posts, we are confident that the established criteria are robust enough to classify the remaining posts.
Overall, manual analysis of 1,860 posts from 8 \textsc{StEx} sites yielded 62 fairness posts from 5 \textsc{StEx} sites, as shown in the \textit{2nd Extraction} of Table \ref{tab:extraction}. In total, we manually identified 136 fairness posts from a total of 4,178 posts and used these posts to answer our research questions. Our small dataset aligns with common qualitative research practices. For instance, in another study on SO, researchers found 315 relevant comments out of 2,000 sample comments \cite{khristul2022}. In a different study about test smells, the authors extracted 101 relevant posts from a pool of over 8.4 million posts \cite{martins2023hearing}. 

\section{Findings}
\subsection*{RQ1:  How often is fairness, relating to human and social factors in SE, discussed on \textsc{StEx} sites?}
To investigate the proportion of fairness-related questions and answers in human and social factors in SE, as well as how frequently the topic is discussed, we began by counting the number of fairness questions and fairness answers. We calculated the proportion of fairness posts related to human and social factors in SE, as well as candidate posts. Candidate posts refer to all posts that use the word \textit{fairness} and related terms as detailed in Section \ref{methodology}. Our aim was to determine how many of the candidate posts specifically address fairness in the human and social aspects of SE.

\underline{\textbf{Results}}. 
The 136 fairness posts comprise of 28 (20.59\%) questions related to fairness, and 108 (79.41\%) fairness answers. These 108 answers correspond to a total of 96 unique questions (which we refer to as \textit{corresponding questions}). Table RQ1 
depicts the breakdown of each of the five \textsc{StEx} sites' question/answer proportion. 
No posts were found from three other \textsc{StEx sites}. We can see that the majority of fairness posts (87 posts) come from \textsc{WP StEx}, while the least (1 post) from \textsc{DO StEx}. 
Since we analyzed a random sub sample of the posts, we have no reasons to believe how the distribution of post population would differ from sample.


\begin{table}[]
\captionsetup{font=footnotesize}
\caption*{Table RQ1: Proportion of Fairness Posts. \textit{Fairness Posts} refers to the total number of posts related to fairness, which is the sum of \textit{Fairness Questions} and \textit{Fairness Answers}. \textit{Percentage} represents the proportion of \textit{Fairness Posts} compared to the posts generated
from the filtering steps, i.e., \textit{Candidate Posts}.}
\label{totalpost}
\footnotesize
\begin{tabular}{lrrrrr}
\begin{tabular}[l]{@{}l@{}}\textsc{StEx}\\Sites\end{tabular} & \begin{tabular}[l]{@{}l@{}}Fairness \\ Questions\end{tabular} & \begin{tabular}[l]{@{}l@{}}Fairness \\ Answers\end{tabular} & \begin{tabular}[l]{@{}l@{}}Fairness \\ Posts\end{tabular} & \begin{tabular}[l]{@{}l@{}}Candidate Posts\end{tabular} & Percentage      \\ \hline
\textsc{WP}                 & 13                 & 74               & \textbf{87}          & 598            & \textbf{14.5\%} \\
\textsc{OS}               & 4                  & 4                & 8                    & 59             & 13.6\%          \\
\textsc{SE}       & 4                  & 18               & 22                   & 369            & 6.0\%           \\
\textsc{PM}         & 6                  & 12               & 18                   & 301            & 6.0\%           \\
\textsc{DO}                    & 1                  & 0                & \textbf{1}           & 27             & \textbf{3.7\%}  \\ \hline
\textbf{Total}                           & 28                 & 108              & 136                  & 1,354          & 10.0\%      \\ \hline
\end{tabular}
\end{table}
\textsc{WP \textsc{StEx}} also has the highest percentage (14.5\%) of fairness in SE posts. This implies that around 14 in every 100 candidate posts on \textsc{WP \textsc{StEx}} discuss human and social aspects of SE. On \textsc{DO \textsc{StEx}}, on the other hand, around 3 (3.7\%) in every 100 candidate posts involve fairness problems. Finally, 136 (10.0\%) of the 1,354 candidate posts across all 5 \textsc{StEx} sites discuss about fairness in SE. This suggests that there are many types of fairness discussions in these \textsc{StEx} sites, but 10 in every 100 candidate posts discuss fairness in human and social aspects of SE. This is in line with the finding of the closest study by Krishtul et al. \cite{khristul2022} that around every 15 in 100 comments in \textsc{SO} violate human values. 


\subsection*{RQ2: What characterizes fairness discussions on \textsc{StEx} sites?}
\subsubsection*{\textbf{RQ2.1: What are the contexts in which fairness is discussed?}} \label{rq2.1}
Understanding the context can help in identifying the background situations in which fairness problems are most likely to occur. For instance, if a fairness problem is frequently discussed in a context, such as salary or recruitment, it may suggest systemic issues. 

To answer RQ2.1, the first author used an open coding approach \cite{glaser_strauss_strutzel_1968} to determine and categorize the context of an event involving fairness concerns (e.g., customer/client relations, recruitment, treatment, and so on). This method entailed reading fairness posts from one \textsc{StEx} site, defining contexts that would cover the posts, rereading the posts, and applying the contexts. We defined the contexts by considering the background situations or factors surrounding a fairness post. For example, we looked at the subject matter (e.g. performance review, job interview) or the subjects being discussed (e.g. developers, project manager). The author then read the fairness posts from a second \textsc{StEx} site and used the contexts generated for the previous \textsc{\textsc{StEx}} site. In cases where the existing contexts did not match with any of the posts from the second \textsc{\textsc{StEx}} site or new contexts were required, we generated new contexts. We then repeated the process on the remaining posts from the remaining \textsc{\textsc{StEx}} sites.

The approach was used to categorize 28 fairness questions and 108 fairness answers. When applying the approach on to the fairness answers, we referred to the 96 corresponding questions to obtain a better sense of the overall context. The outcomes of the open coding process were shared with the second and third authors. For validation, 
the three authors discussed each context and its examples, but no new results emerged.

After obtaining the contexts, we examined the most popular ones by using three distinct complementing measures of popularity that have been used in previous work: average number of views, average number of posts marked as favourite, and average score \cite{Ahmed2018, Bagherzadeh2019}. Average views indicate the interest of the community by presenting how often a post is viewed by both visitors and the site's registered users \cite{Ahmed2018}. In \textsc{SO}, a high view count indicates that more developers face the same or a similar problem \cite{Nadi2016}. Average number of favourites refer to the degree of interest in the solution to the problem addressed in the question \cite{Bajaj2014}, while average scores measure the importance of the topic to the community \cite{Bajaj2014}. Because view counts, favourite counts, and scores are unique to questions only, we only considered 28 fairness questions and 96 corresponding questions. We recognized that the previous related studies \cite{Ahmed2018, Bagherzadeh2019, Bajaj2014, Nadi2016} examined \textsc{SO}. However, \textsc{StEx} is a network of sites, with \textsc{SO} being one of them. All \textsc{StEx} sites utilize the same software and process, but are generally oriented towards different communities. Therefore, it is appropriate to compare these measures across different \textsc{StEx} sites. 

We also discovered that there was 60\% missing data in the favorite count, therefore we eliminated favorite count from the measurement. We further found that the distributions of view counts and score were skewed to the right (skewness 2.78 \& 3.14 respectively). Therefore, we calculated the measures using median instead of the mean \cite{MacGill1981}.



\underline{\textbf{Results}}. 
We identified 10 contexts in which fairness problems occurs, according to fairness posts in \textsc{StEx} sites. These are \textit{income}, \textit{treatment}, \textit{allocation of work}, \textit{recruitment}, \textit{working hours}, \textit{demand}, \textit{evaluation}, \textit{authorship}, \textit{policy}, and \textit{customer/client relations} (see Figure RQ2.1). Out of all the contexts, most posts discuss \textit{income} (24), while only a few of the posts discuss \textit{customer/client relations} (5). This finding indicates that practitioners primarily ask and receive responses regarding the outcome they earn as a result of the work they have accomplished. 
As can be seen in the Figure RQ2.1, fairness posts from \textsc{WP \textsc{StEx}} cover 8 out of 10 fairness contexts. This finding suggests that \textsc{WP \textsc{StEx}} is an ideal resource for understanding practitioners' perspectives on fairness in a variety of contexts, particularly \textit{income}, \textit{treatment}, and \textit{working hours}. Closer inspection of the figure shows that, if we want to understand more about fairness in \textit{policy}, \textsc{WP \textsc{StEx}} is the most suitable place to start. If fairness in \textit{authorship} is something we are concerned about, then \textsc{OS \textsc{StEx}} is the most recommended site to go. 
\begin{figure}[]
\centering
\includegraphics[width=0.38\textwidth]{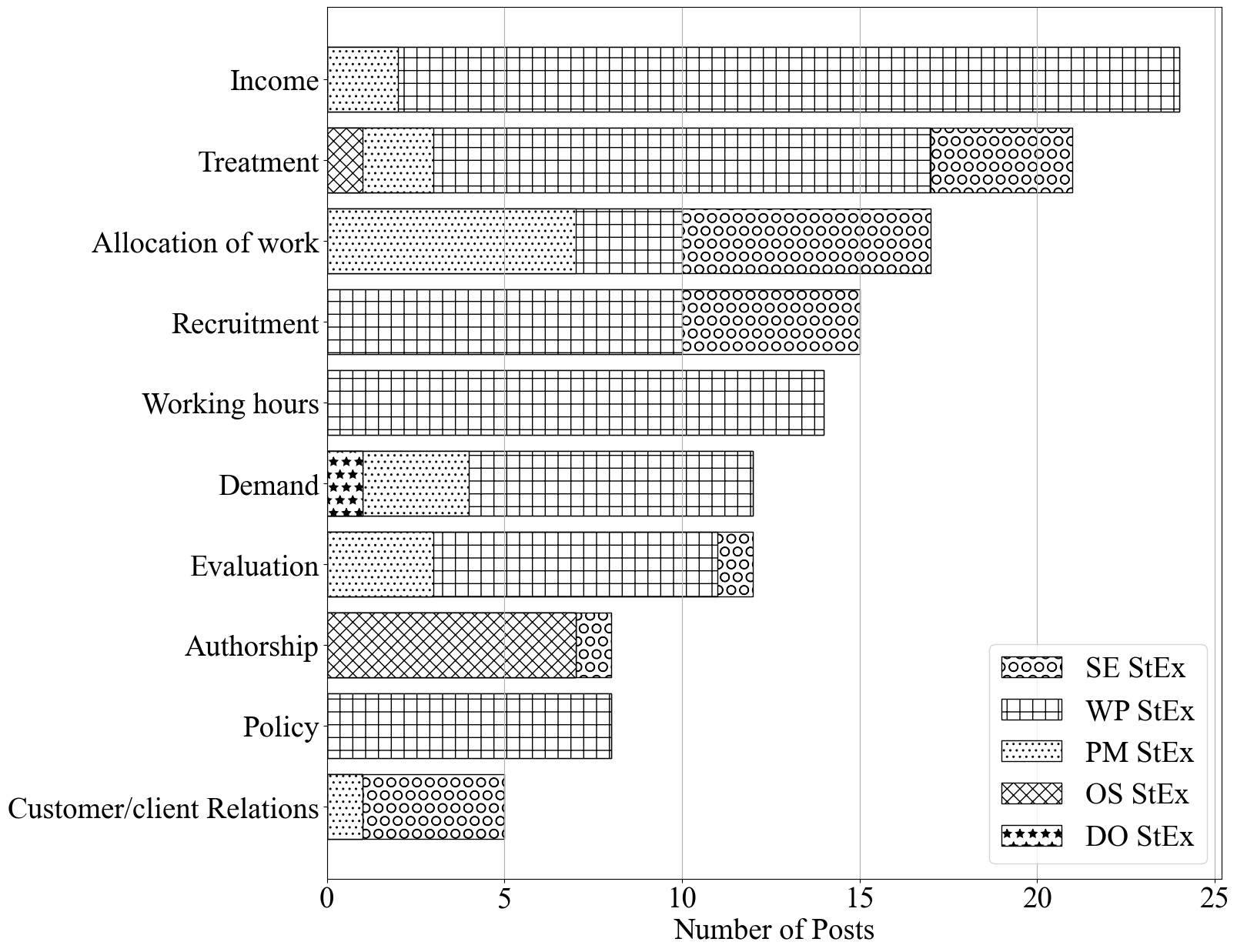}
\captionsetup{font=footnotesize}
\caption*{Figure RQ2.1: Contexts of Fairness Posts}
\label{fig:contexts}
\end{figure}
\begin{table*}[]
\footnotesize
\captionsetup{font=footnotesize}
\caption*{Table RQ2.1: Characterization of Fairness Posts. 
This includes the ten \textit{Fairness Contexts}, their \textit{Definition}, the total number of posts per context (\textit{N}), and their popularity metrics (\textit{Median View} and \textit{Median Score}). For every context, \textit{Example Post} is provided. The last two columns show the \textit{Fairness Dimension(s)} and \textit{Protected Attribute(s)} (RQ2.2) with respect to the \textit{Example Post}. Due to space limit, complete characterization of fairness posts is made available in our repository \cite{figshare}.}
\label{tab:char}
\begin{tabular}{p{.09\linewidth}p{.2\linewidth}rm{2.5em}m{2.5em}p{.325\linewidth}m{5em}m{5em}}
\multirow{2}{.09\linewidth}{Fairness Context} & \multirow{2}{*}{Definition}                                                                               & \multirow{2}{*}{N} & \multirow{2}{3em}{Median View} & \multirow{2}{3em}{Median Score}  & \multirow{2}{*}{Example Post} & \multirow{2}{6em}{Fairness Dimension(s)} & \multirow{2}{6em}{Protected Attribute(s)} \\ \\ \hline
\multirow{3}{*}{Income}                            & Fairness of the outcome that a software developer/engineer earns from the work they have accomplished       & \multirow{3}{*}{\textbf{24}}                                     & \multirow{3}{*}{3734.0}               & \multirow{3}{*}{\textbf{3.0}}                   & \say{Lots of people are quick to point other reasons why you may be getting paid less, however the most likely explanation for most of the difference is gender,} \cite{salary}                   & \multirow{3}{7em}{Distributive \& Procedural}            & \multirow{3}{*}{Gender}                                  \\ \hline
\multirow{4}{*}{Treatment}                         & \multirow{4}{15em}{Fairness of how software developers/practitioners interact with or treat someone}             & \multirow{4}{*}{21}                            & \multirow{4}{*}{4171.0}                        & \multirow{4}{*}{12.0}                           & \say{Are most programmers younger because younger programmers tend to get hired or is this a profession where the only way to get a promotion is to get a job doing something else?} \cite{treatment}                    & \multirow{4}{*}{Procedural}                             & \multirow{4}{*}{Age}                                     \\ \hline
\multirow{4}{.5\linewidth}{Allocation of Work}                & \multirow{4}{15em}{Fairness of the process of arranging resources and labor in a SE project or task}                         & \multirow{4}{*}{17}                                     & \multirow{4}{*}{\textbf{918.0}}                & \multirow{4}{*}{6.0}                            & \say{Generally, he just waits the whole day for phone calls which stay absent not too rarely. His bug reports usually contain only screenshot of an error code or -message sent by email.} \cite{allocation}                       & \multirow{4}{*}{Procedural}                             & \multirow{4}{*}{-}                                       \\ \hline
\multirow{3}{*}{Recruitment}                       & \multirow{3}{15em}{Fairness of the process of hiring software developers/practitioners for an organization}                   & \multirow{3}{*}{15}                                     & \multirow{3}{*}{\textbf{6596.0}}                        & \multirow{3}{*}{\textbf{20.0}}                  & \say{So, I already want to hire more people of color, more women, etc. The frustration I've had—especially on small teams in the midwest—is how to actually do that?}   \cite{recruitment}                     & \multirow{3}{*}{Procedural}                             & \multirow{3}{7em}{Gender, Race, Colour}                    \\ \hline
\multirow{3}{.1\linewidth}{Working Hours}                     & \multirow{3}{15em}{Fairness of the amount of time a software developer/engineer spends doing work during a day}               & \multirow{3}{*}{14}                                     & \multirow{3}{*}{3093.0}                        & \multirow{3}{*}{13.5}                           & \say{Is it fair I'm gonna spend my time on issues other the specific work for the client and possibly won't get compensated for it ?} \cite{workinghours}                    & \multirow{3}{*}{Distributive}                           & \multirow{3}{*}{-}                                       \\ \hline
\multirow{3}{*}{Demand}                            & \multirow{3}{15em}{Fairness of insisting on something that someone with superior authority demands}                           & \multirow{3}{*}{12}                                     & \multirow{3}{*}{1057.0}                        & \multirow{3}{*}{7.0}                            & \say{Assuming these demands weren't made retroactively, is 7.5\% equity stake for Bob's skills and one time work a reasonable request?} \cite{demand}                      & \multirow{3}{7em}{Distributive \& Informational}         & \multirow{3}{*}{-}                                       \\ \hline
\multirow{3}{*}{Evaluation}                        & \multirow{3}{15em}{Fairness of the process of assessing or estimating the quality, importance, amount, or value of something} & \multirow{4}{*}{12}                                     & \multirow{3}{*}{1646.0}                        & \multirow{3}{*}{3.5}                            & \say{Putin isn't very sympathetic to US companies and will try to impose financial and even criminal sanctions on the companies that \say{cut ties} with their Russian workers} \cite{evaluation}                   & \multirow{3}{*}{Distributive}                           & \multirow{3}{*}{Ethnicity}                               \\ \hline
\multirow{3}{*}{Authorship}                        & \multirow{3}{15em}{Fairness of crediting contributors for their contributions}                                                & \multirow{3}{*}{8}                                      & \multirow{3}{*}{946.5}                         & \multirow{3}{*}{7.0}                            & \say{Now that other developers are also contributing to the project how would you suggest donations are fairly distributed?} \cite{authorship}                       & \multirow{3}{*}{Distributive}                           & \multirow{3}{*}{-}                                       \\ \hline
\multirow{3}{*}{Policy}                            & \multirow{3}{15em}{Fairness of a set of ideas or a plan of what to do in particular situations that has impact on SE}         & \multirow{3}{*}{8}                                      & \multirow{3}{*}{2168.5}                        & \multirow{3}{*}{19.5}                  & \say{I'm trying to create a developer-centric compensation policy.  Which is to say it focuses heavily upon transparency, flexibility, and fairness.} \cite{policy}                       & \multirow{3}{7em}{Distributive \& Procedural}            & \multirow{3}{*}{-}                                       \\ \hline
\multirow{4}{.1\linewidth}{Customer/client Relations}         & \multirow{4}{15em}{Fairness in the relationship between customers/clients and software practitioners or software companies}   & \multirow{4}{*}{\textbf{5}}                                      & \multirow{4}{*}{3390.0}                        & \multirow{4}{*}{14.0}                            & \say{We gave our software to an escrow agent, and we believe that is fair. But giving them our code seams to us unfair, knowing that customer will then go on to develop based upon our code.} \cite{ccrelations}                       & \multirow{4}{*}{Distributive}                           & \multirow{4}{*}{-}                                       \\ \hline
\end{tabular}
\end{table*}

Table RQ2.1 provides \textit{median views} and \textit{scores} for 10 fairness contexts. We observe that posts about fairness in \textit{recruitment} has the highest \textit{median view}, suggesting frequent user engagement. In contrast, fairness in \textit{allocation of work} posts has the lowest \textit{median view}, indicating lowest user engagement.
Additionally, posts about fairness in \textit{recruitment} often receive higher \textit{median scores}, implying that they are well-received and considered valuable by the community \cite{Ahmed2018}. 
Surprisingly, the \textit{median score} for fairness in \textit{income} is the lowest. One possible explanation for the low \textit{median score} is the diverse and often controversial nature of discussions related to fairness in \textit{income}. Different users may hold varying opinions on what constitutes fairness in \textit{income}, leading to a wide range of scores for posts on this context.

\subsubsection*{\textbf{RQ2.2: Which dimensions of fairness are discussed?}}
To determine fairness dimensions, we used the Collquit's  questions \cite{Colquitt2001} which are shown in Table \ref{tab:dimensions}, associated with 136 fairness posts. When applying the measurement on to the 108 answers, we referred to the 96 corresponding questions to obtain a better sense of the overall context. 
For each fairness post, the first author assessed which Colquitt's \cite{Colquitt2001} question best represents the fairness problems and took note of the related fairness dimension. For validation, the first, second, and third authors discussed the dimensions identified from the analysis, along with their respective examples.

We also examined the posts to see that individual's protected attributes are mentioned, among those are identified by the Article 21 - Non-discrimination of EU Charter of Fundamental Rights \cite{EUrights_2022} which states that: \say{Any discrimination based on any ground such as sex, race, colour, ethnic or social origin, genetic features, language, religion or belief, political or any other opinion, membership of a national minority, property, birth, disability, age or sexual orientation shall be prohibited.} By identifying whether the post mentions any of these protected attributes, we can identify human factors that may contribute to the occurrence of fairness problems.

\underline{\textbf{Results}}. 
\begin{figure}[]
\centering
\includegraphics[width=0.24\textwidth]{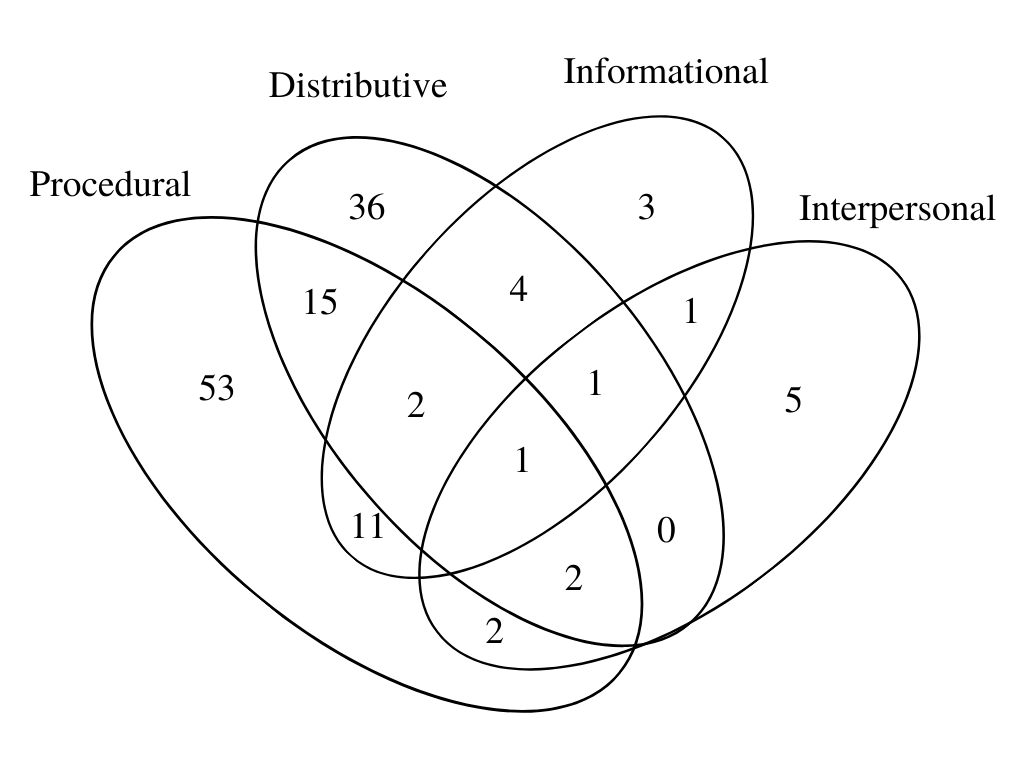}
\captionsetup{font=footnotesize}
\caption*{Figure RQ2.2: Fairness Dimensions Discussed in the Posts}
\label{fig:dimensions}
\end{figure}
We discovered that some posts have only one of the four fairness dimensions, while other posts with a combination of multiple dimensions, as summarized in Figure RQ2.2. With 53 fairness posts (38.97\%), \textit{procedural fairness} is the most discussed fairness dimension. Table RQ2.1 row 2 shows the example post \cite{treatment} which concerns \textit{procedural fairness}. This dimension refers to fairness in decision-making processes that influences an outcome. In the example post \cite{treatment}, the user discusses the possible differences in procedures when age is considered, such as when applying for a job or getting hired. This discussion suggests that the procedure may not be free of bias, which is one of the criteria for \textit{procedural fairness} as can be seen in Table \ref{tab:dimensions}.

There are 25 out of 136 fairness posts that discuss protected attributes, with \textit{gender} (8 posts) being the most frequently mentioned attribute. On the other hand, \textit{age} is referenced in 6 separate fairness posts, \textit{ethnicity} in 5, \textit{religion} and \textit{disability} are mentioned 2 times each, while \textit{language} is mentioned in 1 fairness post. Only 1 fairness post specifies 3 protected characteristics at the same time: \textit{gender}, \textit{race}, and \textit{color}. 
However, there are 111 fairness posts that do not mention any protected attribute. We observe that posts about fairness in human and social aspects on the 5 \textsc{StEx} sites we studied do not always depend on individual’s protected attributes but rather a broader range of factors that can affect fairness.

\subsubsection*{\textbf{RQ2.3: What are the characteristics of the post owners?}}

To answer RQ2.3, we first counted the users associated with fairness post, both with questions and answers. For each post we identified the \texttt{OwnerUserId} and eliminated duplicates. Duplicates indicate that there were users who posted more than one fairness post, which we investigated after. We then ran a query \cite{figshare} on the \texttt{Users} data of each site selecting the gathered distinct \texttt{OwnerUserId} to obtain more publicly available information about each user in order to extract users' location and role.
The query returned 123 rows, of which two had blank \texttt{OwnerUserIds} so we excluded them from subsequent analysis.
We extracted geographical location information by looking at any inferable location from the \texttt{Location}, \texttt{Website}, and \texttt{AboutMe} sections. Then, we categorized the extracted locations into countries, e.g, categorizing Oxford and Manchester to United Kingdom. 

We explored the potential impact of regional differences on the occurrence of fairness problems by inferring a possible relationship between the location of fairness post owners and the geographic distribution of users on each site. We ran a query \cite{figshare} on all 5 sites to obtain the geographic distribution of the users. Given the locations provided is in text free format (i.e., users type their location), we categorized them into countries by using the Bing Maps API \cite{Brundritt_Munk_French_2022} (which helps to get the country name that corresponds to location information provided as a query string). 

It has been observed that unfair situations can be viewed from different perspectives such as that of an employee \cite{gilliland2008justice}, a manager \cite{barclay2005exploring}, or a third party \cite{Skarlicki2015}. Therefore, we also looked for potential differences in fairness posts between roles \textit{managers} and \textit{employees}. We attempted to infer users' roles based on the information written in users' \texttt{Website} and \texttt{AboutMe} sections.

\underline{\textbf{Results.}} 
Our analysis on users' information indicates that there are 121 distinct users. Of these users, 111 posted only one fairness post, while 10 posted more than one. Seven users posted two posts, while three others posted three posts. This finding is likely related to the possibility that, in \textsc{StEx} sites, there is no group of users that exclusively care about fairness in human and social aspects of SE, but rather several individuals with unique fairness concerns and perspectives.

With regards to inferable users' geographical location, 75 (61.98\%) of 121 users make their location available, while 46 others (38.02\%) do not. Among the 75, we inferred 14 distinct users' countries (see our repository \cite{figshare}) with the \textit{United States} accounting for 50.67\% (38 users). 
Comparing this result to the geographic distribution of overall users on 5 sites, the distribution of users writing fairness posts is similar to \textsc{\textsc{StEx}} sites population distribution \cite{figshare}. This interpretation is more likely to apply to \textsc{WP StEx}, \textsc{SE StEx}, and \textsc{PM StEx} where the majority of users posting fairness posts put \textit{United States} as their location. 
However, for the remaining 2 sites, we have insufficient data to make any remarks. 
We found majority of users in both \textsc{OS StEx} and \textsc{DO StEx} put \textit{United States} as their location. Whereas our result for the two sites \cite{figshare} shows that 2 \textsc{OS StEx} users and 1 \textsc{DO StEx} user, who posted fairness posts, put the \textit{United Kingdom} and \textit{Netherlands} as their location, respectively.
Due to the limited data, we cannot make generalizations about the likelihood of users from different countries discussing fairness problems. 

As for the user roles, 52  (42.98\%) of 121 users made information about their roles public, resulting in 20 distinct roles. Among these users, 14 are \textit{Software Developers}, 12 are \textit{Software Engineers}, 6 are \textit{Software Programmers}, 2 are \textit{Product Managers}, and 2 are \textit{Engineering Managers}. Our analysis found 11 fairness posts from \textit{managers} and 47 from \textit{employees}. Check our supplementary material \cite{figshare} for the user roles categorization. Among the posts from \textit{managers}, 
fairness in \textit{working hours} was discussed the most (3 posts, 27.27\%), while \textit{procedural fairness} being the most common fairness dimension (5 posts, 45.45\%). For posts from \textit{employees}, 
\textit{procedural fairness} (22 posts, 46.81\%) being the most common fairness dimension, while \textit{treatment} is mostly discussed (13 posts, 27.66\%).
We observed differences in the most common fairness contexts discussed by \textit{managers} and \textit{employees}.
These differences may suggest that they have distinct approaches to responding to unfairness. Previous research in OJ has found that managers tend to remedy unfairness \cite{gilliland2002emerging}, while employees use various coping strategies to deal with it \cite{aquino2001employees}.

\section{Discussion}
\subsection{Reflection on \textsc{StEx} Sites}
\label{section5.1}

Out of 181 \textsc{StEx} sites, we chose 8 sites that potentially cover a combination of broader and technical aspects of SE. We found that three of them did not have any fairness posts related to human and social aspects of SE. Possible explanation for the lack of fairness-related posts on \textsc{SO} is that the fairness discussion on the site typically revolves around fairness in computation and algorithms, rather than fairness related to human and social aspects. 
This finding may help us understand that even though \textsc{SO} covers problems unique to software development, it is strictly limited to algorithm, code or programming-related topics.
We also did not find a single fairness post on \textsc{CS \textsc{StEx}}. A possible explanation for this might be that the topics in this site are mainly theoretical or practical computer science. 
Similarly, we found little useful information from \textsc{CR \textsc{StEx}} because its posts are mainly dominated by software testing inquiries. This finding does not support what previous study has suggested, that fairness is an essential property of code reviewing \cite{German2018}.  

On the remaining \textsc{StEx} sites we investigated, we observed that the posts were more focused on generic SE, and related to human and social aspects. Researchers and practitioners can actually learn more about how practitioners perceive fairness. 
The contexts in which fairness problems occurs are varied, and each has the potential to improve the established literature on human and social aspects of SE. For example, researchers can investigate what practitioners consider to be fair treatment. \textsc{WP \textsc{StEx}}, \textsc{PM \textsc{StEx}}, and \textsc{SE \textsc{StEx}} are the sites researchers are most likely to find discussions on fair treatment. This research can add to what is known about non-appropriate treatment in code review, such as destructive criticism being rather widespread and perhaps contributing to the lack of gender diversity in the software industry \cite{Gunawardena2022DestructiveCI}.

Our insights might very well also be valuable to software practitioners. For example, understanding what it means to have fair outcome from they work they have accomplished (i.e., income). Practitioners can read the discussions and ask questions on \textsc{WP \textsc{StEx}} and \textsc{PM \textsc{StEx}} to determine how fairness in income is perceived and addressed within the community. This can be a significant contribution to what has already been established, such as link between fair compensation and employee satisfaction \cite{Abdin2020EFFECT}.
Further research might also extend the study to other \textsc{StEx} sites (e.g., \textsc{Software Quality Assurance \& Testing StEx}, \textsc{Freelancing StEx}) or explore other grey literature (e.g., \textsc{Reddit}, \textsc{X}, \textsc{Quora}) that possibly cover more generic SE in identifying posts related to fairness in human and social aspects.

\subsection{Contexts of Fairness Discussions}

In RQ2.1 we found that practitioners mostly discuss about fairness in \textit{income} (e.g., \cite{exampleincome}). This finding suggests that practitioners are primarily focused on understanding and discussing the outcomes they achieve as a result of their work. It reflects an interest in the tangible results and impacts of their efforts. One study by Bhal et al. emphasizes the need for human resource departments to improve pay and raise determination processes to retain software professionals in competitive salary environments \cite{bhal2007}. Another study highlights that gender remains a significant factor in assessing software developers' salaries \cite{Dattero2005Assessing}.


According to the popularity metric, 
fairness \textit{recruitment} (e.g., interviewing someone you know \cite{examplerec1234}) is the context in which many of the user's both visitors and registered users read the discussions and find it valuable. This has the connotation that recruitment is a fundamental aspect of the professional world, and many people face and relate to the same problem and wish to discover the solution. 
It also indicates that posts in these situations meet the needs of practitioners and provide relevant information that they desperately seek.
Previous research  
has 
proposed a list of recommendations to assist companies in enhancing their recruitment practices, including being mindful of the language used during the first contact with the candidate \cite{Behroozi2020}.



\subsection{Fairness Dimensions in SE} 
We observed that most posts on fairness are primarily concerned with the fairness in decision-making processes that influence the outcome (i.e., \textit{pocedural fairness}). These processes include work allocation process that is applied consistently \cite{allocation}, interview procedure \cite{interviewp} and performance review \cite{performancer} that is based on accurate information, and software testing that is free from confirmation bias \cite{cb}. The second most mentioned dimension is \textit{distributive fairness}. For instance, problems in this dimension deal with practitioners feeling underpaid \cite{salary}, 
and giving copyrights to software contributors \cite{copyright}. The high number in these two dimensions is inline with fairness study in code review that found \textit{distributive} and \textit{procedural fairness} (mainly consistency and bias-suppression) to be the most common problems mentioned among OpenStack developers \cite{German2018}.

\textit{Interpersonal fairness} was mentioned in  posts mostly discussing about treatment, such as how a software engineer might not being treated with dignity and respect \cite{respect}, or how unhelpful and heavy criticism in code reviews leads to argument \cite{argument} (combination with \textit{informational fairness}). Whereas, posts about \textit{informational fairness} are concerned with whether information related to a process is reasonable or not, such as demand to spend developer's non-work hours to learn new frameworks \cite{frameworks}, honesty in providing personal information when applying to be a developer \cite{applying}, and justification of decisions about criteria in allocating developers to a new cooler project \cite{coolerp}. It can therefore be assumed that not all fairness is equal, and fairness is highly dependant on the context the problem is taking place, and on individual's values and assumption that come from many aspects of life. This finding may help us to understand that not all fairness can be achieved, but one can understand which fairness matters the most to the practitioners in a given context. 

We also observed fairness posts with a combination of multiple dimensions, such as 15 posts concerning a combination of \textit{procedural} and \textit{distributive fairness}, and 11 posts on \textit{procedural} and \textit{informational fairness}. The former example involves the development of a developer-centric compensation policy (\textit{distributive fairness}) with an emphasis on transparency, flexibility, and an unbiased approach (\textit{procedural fairness}), particularly in the context of performance-based bonuses \cite{intersection1}. The latter example describes a lead programmer creating an evaluation process that brings fairness to the evaluation process, ensuring that decisions about employee performance are based on consistent criteria (\textit{procedural fairness}) and that decisions are effectively communicated
(\textit{informational fairness}) \cite{intersection2}. Such intersectionality highlights the complexity of fairness problems in SE, where different aspects of fairness are intertwined and impact each other.

Finally, posts about fairness in human and social aspects on \textsc{StEx} sites do not always depend on variables such as an individual's protected attributes. It is therefore likely that the feeling of being treated fairly or unfairly is not limited to discrimination along the established protected attributes. Of all the fairness posts mentioning protected attributes, \textit{gender} seems to be most discussed attribute. This result reflects studies 
that found that majority of diversity studies in SE were focused on gender diversity \cite{Menezes2018,Rodriguez-Perez2021}. 
Despite the extensive research on gender diversity, women remain underrepresented in the field, and there is evidence of bias against them in some communities, as well as negative perceptions about women working in software teams \cite{Rodriguez-Perez2021}. This highlights the need for continued consideration and study of gender-related fairness.


\subsection{Implications for Enhancing Fairness}
This study examines how software practitioners experience fairness problem, offering insight that may help researchers in understanding what is important to these practitioners and improving work practices \cite{John2005}.
We have compiled a collection of posts on fairness issues faced by practitioners, including context-specific problems.
The findings could also be explored by researchers to explore how fairness impacts SE outcomes.  
Fairness, as explored in OJ literature, notably impacts job satisfaction \cite{Otaye2014Mapping,Choi2014Organizational,Dubinsky1993Effects, Qureshi2019Impact}, a key driver of productivity in SE \cite{Storey2021}. Story et al. found this relationship is affected by challenges in SE and various workplace variables \cite{Storey2021}. Our research identified fairness issues in SE, which also can lead to further exploration into their importance and resolution, inspired by Begel et al. \cite{Begel2014}. These findings could lead to important questions about which fairness problems are important and how they can be effectively addressed in SE. Our mapping to the OJ literature also creates opportunities to bring insights from other literature into SE. Furthermore, by using the labelled data, an automated tool for identifying and classifying fairness posts could also be developed
, as done in the work of Callan et al. \cite{Callan2022}.

To better understand how fairness is experienced in SE, practitioners should reflect on Colquitt’s fairness theory and consider everyday situations unique to SE, as observed in this study. Colquitt’s theory is based on practical and specific principles \cite{Colquitt2015}, and improving fairness perceptions can enhance organizational outcomes \cite{Colquitt2001}. 
For example, procedural fairness is highly correlated with job satisfaction \cite{Mossholder1998,Wesolowski1997}, and creating a fair environment could lead to more satisfied developers, helping companies attract and retain talent \cite{Storey2021}. To achieve procedural fairness, practitioners should consider offering more consistency, accuracy, and reduced sources of bias \cite{Colquitt2015}.

The identified 136 fairness posts can be interpreted in several ways. Take from the example post \cite{treatment} from Table RQ2.1, that discusses ageism in the software industry, suggesting concerns about unfair treatment based on age. This implies that the user finds ageism in SE to be a problem and this can pinpoint the practitioners on further promoting equal treatment regardless of age. 
Overall, these posts offer a nuanced understanding of the perception and experience of fairness within SE. They illuminate key areas where fairness problems are prevalent. 

\section{Threats to Validity}
Construct Validity refers to how well a measurement captures its intended aspects \cite{Ralph2018}. However, applying Colquitt's fairness measure in RQ2.2 could be influenced by researchers bias. To mitigate this threat, we conducted consensus coding process between 3 authors, and consulted seminal papers in OJ literature \cite{Colquitt2001,Colquitt2002,Colquitt2012, Colquitt2013,Colquitt2015} as well as previous research that applied fairness theory in SE \cite{German2018}. The results of the study was confirmed by OJ researcher, who agreed that procedural fairness is the most crucial fairness dimension, which supports our observation in RQ2.2.

Internal Validity is described as threats that may have influenced our findings \cite{Wohlin2012}. 
We randomly selected 300 posts from each of the 8 \textsc{StEx} sites (1,860 posts total) during the second post filtering, manually extracted fairness posts, and performed a qualitative analysis for RQ2.1. However, our findings may be threatened by several factors. We may have missed crucial posts about fairness in SE due to our random selection of 1,860 posts. The manual extraction of fairness posts may also be subjective and prone to mistakes, although we took steps to reduce this threat by having multiple authors involved in the process. We addressed the potential issue of low numbers of selected posts by filtering and manually extracting the data twice, which improved the reliability of our selection. This careful process not only enhanced our current analysis but also laid the groundwork for creating an automated solution in the future. 
Additionally, the process of categorizing posts into contexts in RQ2.1 may be prone to mistakes, but we validated our work by having 3 authors hold several discussions before finalizing the list of contexts. Lastly, we acknowledge the limitations in identifying fairness posts from 8 \textsc{StEx} sites.
However, we concentrated our data collection efforts on sites that we believed were most likely to provide valuable insights into our research questions. We also made attempts to uncover the possible explanations for why we found that 3 of the sites did not have any fairness posts on Section \ref{section5.1}.

External Validity is the degree to which our findings can be applied to different situations \cite{Wohlin2012}. 
We analyzed 136 fairness posts on 8 \textsc{StEx} sites. Our findings are based only on these posts and may not apply to all posts on the 8 StEx sites nor to overall SE concerns. Additionally, they cannot be generalized to other grey literatures. However, the finding that \textit{procedural} and \textit{distributive fairness} are the two dimensions that practitioners are most concerned about may apply more broadly because they reflect what prior studies have reported \cite{German2018}. In RQ2.1, the interpretation of context depends on the post, and one way to diversify information sources is to consider several \textsc{StEx} sites. Therefore, the list of contexts identified does not necessarily reflect the opinions of software practitioners beyond the current sample. This study also did not identify the opinions of practitioners from minoritized groups who might have had different experiences related to fairness. 

\section{Conclusion and future work}

We conducted a study on fairness in SE by analyzing \textsc{StEx} posts. We examined 4,178 posts from eight \textsc{StEx} sites to find discussions about fairness related to social and human factors. We further explored the contexts where the problem arises and the post owners' characteristics. Using Colquitt's list of questions, we also studied the fairness dimensions that practitioners care about. 

Our findings reveal that 136 posts addressed fairness concerns in processes and interactions related to SE.
The diverse range of posts analyzed has uncovered unique fairness perspectives. By focusing on the qualitative data, we have gained a deeper understanding of the fairness problems experienced by several practitioners in the SE field. In conclusion, this study highlights the significance of considering diverse perspectives and understanding fairness issues in SE practice. By addressing these issues, we can strive towards a more fair SE community. In the future, we intend to expand our exploratory study to observe the solution to the fairness problems and how fairness affects the outcome of SE.


\balance
\bibliographystyle{ACM-Reference-Format}
\bibliography{sample-base}

\end{document}